\documentclass[
    ,final            % use final for the camera ready runs
  ]
  {aipproc}

\layoutstyle{8x11single}

\usepackage{amsmath,amssymb,bbm,slashed}
\usepackage{xspace}

\newcommand{\EFTnopi}{EFT($\slashed{\pi}$)\xspace}
\newcommand{\wave}[3]{\ensuremath{{}^{#1}\mathrm{#2}_{#3}}\xspace}

\newcommand{\twoS}{\wave{2}{S}{\frac{1}{2}}}

\newcommand{\twoPone}{\wave{2}{P}{\frac{1}{2}}}

\newcommand{\MeV}{\,\text{MeV}}

\newcommand{\VS}{\vec{\sigma}}

\newcommand{\LRD}{\stackrel{\leftrightarrow}{D}}

\newcommand{\daR}{g^{(^3 \! S_1-^1 \! P_1)}}
\newcommand{\dbR}{g^{(^1 \! S_0-^3 \! P_0)}_{(\Delta I=0)}}
\newcommand{\dcR}{g^{(^1 \! S_0-^3 \! P_0)}_{(\Delta I=1)}}
\newcommand{\ddR}{g^{(^1 \! S_0-^3 \! P_0)}_{(\Delta I=2)}}
\newcommand{\deR}{g^{(^3 \! S_1-^3 \! P_1)}}

\newcommand{\aSing}{a^{(^1 \! S_0)}}
\newcommand{\aTrip}{a^{(^3 \! S_1)}}
\newcommand{\rSing}{r^{(^1 \! S_0)}}
\newcommand{\rTrip}{r^{(^3 \! S_1)}}

\begin{document}

\title{Hadronic Parity Violation in Effective Field Theory}

\classification{11.10.Ef, 11.30.Er, 13.75.Cs, 21.30.Fe}
\keywords      {effective field theory, hadronic parity violation}

\author{Matthias R.~Schindler}{
  address={Department of Physics and Astronomy, University of South Carolina, Columbia, SC 29208}
}

\begin{abstract}
An effective field theory program to analyze and interpret hadronic parity violation in two-, three-, and few-nucleon systems is described. Observables can be parameterized in terms of five low-energy constants, which have to be determined from experimental input. Results for parity-violating observables in the two- and three-nucleon sectors are presented, including a discussion of the relevance of parity-violating three-nucleon interactions.
\end{abstract}

\maketitle

%%%%%%%%%%%%%%%%%%%%%%%%%%%%%%%%%%%%%%%%%%%%
%% MAINMATTER
%%%%%%%%%%%%%%%%%%%%%%%%%%%%%%%%%%%%%%%%%%%%

\section{Introduction}

The weak interactions between quarks induce a parity-violating (PV) component in nucleon-nucleon interactions, which is suppressed by about $10^{-7}-10^{-6}$ compared to the parity-conserving (PC) part of the interaction (for reviews see, e.g., Refs.~\cite{Adelberger:1985ik,RamseyMusolf:2006dz}). Few-nucleon experiments utilizing polarized neutrons are being performed or planned at the Spallation Neutron Source at Oak Ridge National Laboratory, NIST, and other neutron facilities to map out this weak component of the nuclear force. Due to the non-perturbative nature of QCD at low energies, an understanding of how the weak quark-quark interactions manifest themselves in nucleon interactions remains elusive. Traditionally hadronic parity violation has been described in terms of either low-energy transition amplitudes \cite{Danilov} or, more commonly, a meson-exchange model \cite{Desplanques:1979hn}. Here we report on a systematic and model-independent approach using effective field theory (EFT), which includes theoretical error estimates and the consistent treatment of two- and few-nucleon interactions as well as external currents.

\section{Effective field theory}

In the EFT approach, the effective Lagrangian contains all operators consistent with the assumed symmetries of the underlying theory. The operators are organized according to a power counting which predicts the relative size of contributions to physical observables. Each operator is accompanied by a low-energy constant (LEC), which encapsulates the unresolved high-energy physics. The LECs cannot be predicted from the EFT and are commonly extracted from data. Once a set of LECs is determined, it can be used to predict physical observables. 

The energies of current and planned few-nucleon experiments to study PV nucleon-nucleon interactions are low enough that pion exchange cannot be resolved explicitly. It is therefore convenient to use a pionless EFT, \EFTnopi, which only contains nucleons as dynamical degrees of freedom (see, e.g., Ref.~\cite{Bedaque:2002mn}). Using auxiliary dibaryon fields $t_i$ and $s_a$, the leading-order (LO) Lagrangian can be written in terms of five partial wave transition operators \cite{Zhu:2004vw,Girlanda:2008ts,Phillips:2008hn},\footnote{These operators are the field theory equivalent of the Danilov amplitudes \cite{Danilov}.}
\begin{align}\label{Lag:PV}
\mathcal{L}_{PV}^{d}=  -  & \left[ \daR t_i^\dagger \left(N^T \sigma_2  \tau_2 i\LRD_i N\right) \right. +\dbR s_a^\dagger  
\left(N^T\sigma_2 \ \VS \cdot \tau_2 \tau_a i\LRD  N\right) \notag\\
&+\dcR \ \epsilon^{3ab} \, (s^a)^\dagger 
\left(N^T \sigma_2  \ \VS\cdot \tau_2 \tau^b \LRD N\right)  +\ddR \ \mathcal{I}^{ab} \, (s^a)^\dagger 
\left(N^T \sigma_2 \ \VS\cdot \tau_2 \tau^b i \LRD N\right) \notag\\
& \left. +\deR \ \epsilon^{ijk} \, (t^i)^\dagger 
\left(N^T \sigma_2 \sigma^k \tau_2 \tau_3 \LRD{}^{\!j} N\right) \right] + h.c.\ ,
\end{align}
where $a\, \mathcal{O}\LRD b = a\,\mathcal{O}\vec D b - (\vec D a)\mathcal{O} b$, $\mathcal{O}$ is a spin-isospin operator, and 
$\mathcal{I}=\text{diag}(1,1,-2)$.

\section{Two-nucleon sector}

The PV interactions result in a nonzero longitudinal asymmetry $A_L$ for $\vec{N}N$ scattering. Neglecting Coulomb effects, the LO result for the $\vec{p}p$ case is \cite{Phillips:2008hn}\footnote{Results for $np$ and $nn$ scattering are also given in Ref.~\cite{Phillips:2008hn}, but no measurements have been performed to date.}
\begin{equation}
A_L^{pp} = \frac{\sigma_+ - \sigma_-}{\sigma_+ + \sigma_-} = 4pM \sqrt{\frac{\rSing}{\pi}}\,\left[ \dbR + \dcR + \ddR \right]\;, 
\end{equation}
where $\sigma_\pm$ is the total cross section for a proton with helicity $\pm$, $p$ the center-of-mass momentum, $M$ the nucleon mass, and $\rSing$ the effective range in the singlet channel. We have used relations given in Ref.~\cite{Schindler:2009wd} to rewrite the result of Ref.~\cite{Phillips:2008hn} in terms of the LECs $g^{(X-Y)}$ of Eq.~\eqref{Lag:PV}. A measurement at $E=13.6\,\MeV$ \cite{Eversheim:1991tg} can be used to extract the shown combinations of LECs. Coulomb effects can also be taken into account in the EFT framework, but are found to be negligible at the considered energies \cite{Phillips:2008hn}.
 
The PV low-energy $\vec{n}p$ forward scattering amplitude is related to the spin rotation angle of a transversely-polarized neutron beam traveling through a hydrogen target. 
The rotation angle is proportional to a linear combination of the LECs $\daR$, $\dbR$, $\ddR$, and $\deR$. For details see a forthcoming publication \cite{spinrot}. Using an order-of-magnitude estimate for the LECs $g^{(X-Y)}$, the rotation angle for a target density $\rho\approx10^{23}\;\text{cm}^{-3}$ is \cite{spinrot}
\begin{equation}
\label{npestimate}
  \left|\frac{\text{d} \phi_\text{PV}^{np}}{\text{d} l}\right|\approx \left(10^{-7} - 10^{-6}\right)
  \frac{\text{rad}}{\text{m}} \;.
\end{equation}

Information on two independent combinations of LECs can be obtained from two observables in $np\leftrightarrow d\gamma$\,: the photon asymmetry $A_\gamma$ in polarized neutron capture, $\vec{n}p\to d\gamma$, and the circular photon polarization $P_\gamma$ in unpolarized capture, $np\to d\vec{\gamma}$. The photon asymmetry $A_\gamma$ at LO is \cite{Savage:2000iv,Schindler:2009wd,Shin:2009hi}
\begin{equation}
A_\gamma=2M^2\sqrt{\frac{\rTrip}{\pi}}\,\frac{1-\frac{\gamma\aTrip}{3}}{\kappa_1\left( 1-\gamma\aSing \right)}\,\deR,
\end{equation}
while the LO result for the circular photon polarization $P_\gamma$ is \cite{Schindler:2009wd}
\begin{align}
P_\gamma = & - 2\sqrt{\frac{\rTrip}{\pi}}\,\frac{M^2}{\kappa_1\left(1-\gamma\aSing\right)} 
\left[ \left(1-\frac{2}{3}\gamma\aSing\right)\daR + \frac{\gamma\aSing}{3}\sqrt{\frac{\rSing}{\rTrip}}\left(\dbR-2\ddR\right)\right].
\end{align}
Here, $\kappa_1$ is the isovector nucleon magnetic moment, $\gamma$ the deuteron binding momentum, and $a^{(X)}$ ($r^{(X)}$) the scattering length (effective range) in partial wave $X$.

\section{Three-nucleon sector}

In \EFTnopi, the straightforward application of the power counting predicts that three- and few-nucleon interactions are suppressed relative to the leading two-nucleon interactions. However, in the PC sector the solution of the neutron-deuteron scattering equation in the \twoS channel at LO shows a strong dependence on the cutoff $\Lambda$ used to regularize the integral equation \cite{Bedaque:1999ve}. This cutoff dependence can be removed by promotion of a three-nucleon interaction (3NI) to leading order. A similar enhancement of a 3NI in the PV sector would complicate the EFT program, as additional experimental input is required to determine the strength of the interaction.  An analysis of the loop diagrams appearing in the equations for PV $Nd$ scattering shows that no PV 3NI is required at LO and NLO \cite{Griesshammer:2010nd}. The lowest-order PV three-nucleon operators containing one derivative correspond to $\twoS-\twoPone$ transitions with $\Delta I=0,1$. Their spin-isospin structure is different from any potential divergences that might appear up to NLO, and these operators are therefore not required to renormalize the $Nd$ scattering amplitude at this order. This ensures that for calculations with an accuracy of $\approx 10\%$ the five operators of the Lagrangian of Eq.~\eqref{Lag:PV} are sufficient to encode PV effects, and the values of the corresponding LECs can be extracted from three-nucleon experiments.

In Ref.~\cite{spinrot} the spin rotation angle at NLO for a deuteron target is determined from $\vec{n}d$ forward scattering using \EFTnopi consistently for all interactions. The result provides a constraint on a linear combination of the LECs $\daR$, $\dbR$, $\dcR$, and $\deR$. 
An analysis of theoretical errors leads to an estimate of higher-order contributions of roughly $10\%$ as expected. The numerical calculations also confirm the findings of Ref.~\cite{Griesshammer:2010nd} that no PV 3NI is required for renormalization up to NLO. 
A rough estimate analogous to the one for $np$ spin rotation leads to
\begin{equation}   
\label{ndestimate}
\left|\frac{\text{d} \phi_\text{PV}^{nd}}{\text{d} l}\right|\approx   \left(10^{-7} - 10^{-6}\right)\;\frac{\text{rad}}{\text{m}} \;, 
\end{equation}
which is of the same size as the rotation angle estimate in hydrogen of Eq.~\eqref{npestimate} as well as the result of the hybrid calculation in Ref.~\cite{Song:2010sz}.

\section{Conclusions}

The results presented here form part of a comprehensive study of hadronic parity violation in few-nucleon systems. Using \EFTnopi, parity-conserving and parity-violating interactions are treated in a unified framework. The existence of a small expansion parameter not only allows the systematic improvement of calculations, but also the reliable estimation of theoretical errors.  In the \EFTnopi formalism, low-energy PV processes are parameterized in terms of five LECs. The results presented here in principle allow the extraction of these LECs. Future calculations in the three- and few-nucleon sectors then represent predictions of the corresponding observables. At the present, however, not enough experimental information is available from the two- and three-nucleon sectors. An ongoing experimental effort, in particular at the Spallation Neutron Source at Oak Ridge National Laboratory and at NIST, is aimed at gaining improved constraints on PV observables in few-nucleon systems.

%%%%%%%%%%%%%%%%%%%%%%%%%%%%%%%%%%%%%%%%%%%%%%%%
%% BACKMATTER
%%%%%%%%%%%%%%%%%%%%%%%%%%%%%%%%%%%%%%%%%%%%%%%%

\begin{theacknowledgments}

I thank H.~W.~Grie{\ss}hammer, D.~R.~Phillips, and R.~P.~Springer for productive and enjoyable collaborations.
  
\end{theacknowledgments}

\bibliographystyle{aipproc}

\end{document}